**Reproduction of Simulated Acoustics Scenes Using Acoustics Room Transformer for speech intelligibility tests in Clinics**

Ali Fallah[1], Kirsten C Wagener[2], Pascal Pflüger[1], Steven van de Par[1]

[1] Department of Medical Physics and Acoustics, Carl von Ossietzky University of Oldenburg, Cluster of Excellence – Hearing4all, Oldenburg, Germany

[2]Hörzentrum Oldenburg GmbH and Cluster of Excellence – Hearing4all, Oldenburg, Germany

## Abstract:

An Acoustic Room Transformation (ART) method is proposed for clinical applications to render simulated acoustic environments using six loudspeakers in a reverberant room. In typical clinical settings, testing rooms are not acoustically treated and only a limited number of loudspeakers are available. The ART approach enables control of the playback-room reverberation, allowing speech intelligibility tests to be performed under realistic and perceptually compensated conditions. The method focuses on the two most common Speech Reception Threshold (SRT) configurations: $S_0N_0$ and $S_0N_{90}$. One front and one side loudspeaker reproduce the direct sound components, while the remaining four loudspeakers (two in front and two behind the listener) generate the reverberant sound field. The direct sound is optimized in terms of energy, and the reverberant field is adjusted based on both energy and interaural coherence. An additional direct-sound gain is applied to match the clarity index ($C_{50}$) of the reproduced signals to that of the reference target. Results from speech intelligibility tests demonstrate that, with ART processing, the reproduced scenarios achieve SRT values very close to those of the reference signal in both $S_0N_0$ and $S_0N_{90}$ conditions and therefore allow for comparable speech test results independent of the particular test room acoustics.

## I. INTRODUCTION

Speech intelligibility is a central outcome measure in audiology and auditory perception research and is routinely assessed in both clinical and experimental contexts. Typical measurements rely on speech-in-noise paradigms, in which listeners repeat speech presented in background noise under controlled conditions to quantify performance in adverse listening situations. A commonly used metric derived from speech-in-noise testing is the Speech Reception Threshold (SRT), defined as the signal-to-noise ratio (SNR) at which a listener correctly repeats a fixed proportion of the speech material. By convention, the SRT usually refers to the 50% correct point, although other target levels are also used, such as 70%, commonly denoted as SRT70. For optimal experimental control and reproducibility, intelligibility measurements are often conducted using headphone-based presentations, which allow precise control of acoustic conditions. However, headphone testing is not suitable for all applications, for example when assessing listeners using hearing aids or cochlear implants, or when spatial sound reproduction is required. In such cases, loudspeaker-based setups are necessary, albeit with increased influence of room acoustics and reverberation. For example, Winkler *et al.*, 2025 showed that in loudspeaker-based SRT testing, the listening room acoustics can significantly influence SRTs, even in audiometric test rooms with low reverberation times. They therefore recommended that SRT results should be reported together with the test-room characteristics and the loudspeaker configuration.

Recently, speech intelligibility research has expanded beyond laboratory environments through the development of automated and home-based SRT assessments (Ooster *et al.*, 2020) using adaptive procedures and consumer audio devices. These approaches reflect a shift toward intelligibility measurements that better balance experimental control with ecological validity. In this work we focus on speech intelligibility assessment in clinical settings, that incorporate some level of reverberation to allow to assess how well a hearing-impaired listener is able to cope with reverberation. For this, it will be important to be able to electro-acoustically control the room acoustics of the clinical environment to allow comparability across different clinics.

Reverberation has a major influence on speech intelligibility, as it modifies the temporal, spectral, and spatial characteristics of sound reaching the listener. It reduces the clarity of speech by smearing temporal envelopes, masking phonetic cues, and diminishing spatial separation between sources (Haeussler and van de Par, 2019). Haeussler and van de Par (2019) demonstrated that reverberation significantly impairs speech perception and that even small changes in coloration and reverberation can alter perceived speech quality and intelligibility, indicating that listeners are highly sensitive to deviations from reference conditions. Kates *et al.* (2025) conducted a comprehensive study examining the combined effects of noise, reverberation, spatial separation, and signal processing on speech intelligibility. Their results showed that binaural unmasking (the $S_0N_{90}$ advantage) strongly depends on both reverberation and listener group, showing a reduction in spatial benefits as reverberation increases. Specifically, normal-hearing listeners exhibited larger spatial advantages than hearing-impaired listeners, particularly under low-reverberation conditions. Similar findings have been reported by (Reinhart and Souza, 2016; Puglisi *et al.*, 2021; Cueille *et al.*, 2022), who observed that even moderate reverberation leads to measurable reductions in intelligibility, particularly for listeners with hearing impairment. Dong and Lee (2018) demonstrated that reverberation effectively reduces the available signal-to-noise ratio by adding delayed reflections that increase the background energy level, thereby masking important speech cues. Spatial separation between speech and noise sources can partially mitigate this effect, emphasizing spatial configuration as a critical factor for intelligibility in complex acoustic environments (Puglisi *et al.*, 2021). In a comprehensive speech intelligibility modeling study, Rennies *et al.* (2011) examined experimental data and multiple intelligibility models across different reverberation conditions. They demonstrated that spatial release from masking (SRM), the intelligibility benefit of $S_0N_{90}$ relative to $S_0N_0$, is substantially reduced in reverberant environments because late reflections degrade binaural cue fidelity and diminish the head-shadow advantage. Their analysis linked speech intelligibility to key room-acoustic parameters, including the modulation transfer function (MTF) and clarity-based metrics such as C50 and definition ($D_{te}$), which separate the reverberant response into its useful and detrimental (U/D) parts.

Simulated virtual acoustic environments have become an essential tool for studying speech intelligibility under controlled yet realistic reverberant conditions. Headphone-based simulations using Binaural Room Impulse Responses (BRIRs) provide excellent acoustic precision but offer limited ecological validity because they restrict head movement and prevent the use of hearing aids. Measurements using loudspeaker-based systems provide a more representative basis for clinical and perceptual testing. Mansour *et al.* (2021) demonstrated that speech understanding in realistic virtual environments closely matches listener performance in real reverberant rooms, showing that accurate spatial cues such as reflection direction and timing are crucial for perceptually valid simulations. Schütze *et al.* (2025) used the RAZR framework (Wendt *et al.*, 2014) for real-time simulation of perceptually plausible virtual rooms and showed that simulated BRIRs in living-room conditions can closely approximate real-room acoustics. Other studies, such as (Guastamacchia *et al.*, 2024), have extended this research by integrating audio-visual cues to evaluate speech intelligibility in realistic multisensory contexts. Together, these approaches highlight a growing trend toward immersive virtual simulations that reflect everyday listening scenarios, although aligning spatial and visual cues still presents perceptual and technical challenges. Many existing loudspeaker-based systems for studying speech intelligibility employ large multichannel arrays operated in ideal or near-ideal acoustic conditions. For example, Ahrens et al. (2019) used a 64-loudspeaker array to measure and model intelligibility in real and simulated rooms. Xia et al., 2018 investigated the combined effects of noise and reverberation using a multichannel loudspeaker array in controlled acoustic conditions. Badajoz-Davila et al. (2020) employed a three-dimensional loudspeaker array to evaluate cochlear implant users' performance in realistic sound fields, and Gerken et al. (2024) compared 2D and 3D multichannel loudspeaker rendering methods to assess perceptual and technical accuracy.

In practical clinical environments, however, conditions for rendering virtual acoustic environments are rarely achievable. Anechoic rooms and large multichannel loudspeaker systems are typically unavailable due to space limitations, cost constraints, and installation complexity. Consequently, there is a strong demand for simpler and more compact reproduction systems that can deliver perceptually valid acoustic scenes in standard clinical or laboratory rooms without extensive acoustic treatment. Loudspeaker-based room compensation techniques aim to reproduce realistic sound fields by correcting for the acoustic effects of both the loudspeakers and the playback room. The Loudspeaker-based Room Auralization (LoRA) system (Favrot and Buchholz, 2010) employs multiple calibrated loudspeakers to simulate direct and reflected sound within a virtual space, using equalization to correct for loudspeaker directionality, distance, and room response. (Hamdan and Fletcher, 2022) proposed a two-loudspeaker crosstalk-cancellation system for realistic binaural reproduction in hearing-assistive applications, offering high accessibility but limited spatial resolution and robustness under reverberant conditions. Also, related studies (Dong and Lee, 2018; Zucatelli and Coelho, 2021) have shown that combining inverse filtering and time-domain enhancement techniques can improve clarity and intelligibility in complex acoustic conditions.

Considering the limitations of existing spatial audio systems and the need for a method that balances spatial realism, cost-efficiency, and clinical practicality, we proposed the Acoustic Room Transformation (ART) system (Fallah *et al.*, 2023) for reproducing virtual acoustic scenes on loudspeakers while compensating for the acoustics of the reproduction environment. The previously proposed ART approach utilizes only four horizontally arranged loudspeakers in a typical echoic room and aims to reproduce the perceptually relevant acoustic cues of a virtual environment while compensating for playback-room effects. In this method, the direct and reverberant components of a sound field are separated and optimized both spectrally and spatially according to perceptual criteria, ensuring that the reproduced sound field perceptually corresponds to the original. This is achieved through time-invariant filtering and cross-mixing operations that align the spectral and spatial characteristics between the recording and playback rooms. Specifically, the method emphasizes the directional spatial properties of the direct sound and the spatial diffuseness of the reverberant field. In our recent study (Fallah *et al.*, 2025), we extensively evaluated the ART system in terms of its perceptual quality and similarity to the simulated reference generated with RAZR (Wendt *et al.*, 2014). In the present work, we extend the ART framework to a six-loudspeaker configuration, adding dedicated channels for direct speech and noise sources to enable the evaluation of binaural speech intelligibility in room environments. This extension provides a robust and flexible platform for controlled Speech Reception Threshold (SRT) testing under realistic and reproducible acoustic conditions.

## II. ART METHOD

### A. BASIC ART

In our previous work (Fallah *et al.*, 2025), we introduced a practical method for reproducing simulated acoustic scenes in a reverberant playback room using only four horizontally arranged loudspeakers. This approach, termed Acoustic Room Transformation (ART), was developed to ensure that the reproduced sound field in the playback room preserves the perceptually relevant spectral and spatial characteristics of the simulated environment while minimizing changes in coloration and spatial distortion caused by the room acoustics. The similarity between the reproduced and reference signals was evaluated in terms of perceptual audio quality, demonstrating that the reproduced scenes closely matched the simulated reference. A comprehensive theoretical description, implementation details, and perceptual evaluation of this method are provided in that reference. In the ART method proposed in (Fallah *et al.*, 2025), the virtual environment is simulated using the RAZR tool (Wendt *et al.*, 2014), which generates both the reference binaural room impulse response ($BRIR_{ref}$) for headphone playback and also a set of virtual acoustic sources (VASs) corresponding to the direct and reverberant sound components to be reproduced over loudspeakers. The $BRIR_{ref}$ is divided into direct and reverberant parts using a separation time $T_{ref}$:

$$BRIR_{ref}[n] = BRIR_{ref,d}[n] + BRIR_{ref,r}[n]. \tag{1}$$

where $BRIR_{ref,d}[n]$and $BRIR_{ref,r}[n]$denote the direct and reverberant parts of the simulated reference BRIR, respectively.

For the reproduction, the VASs of the simulated room are mapped onto four loudspeakers in the playback room using the vector-based amplitude panning (VBAP) (Pulkki, 1997) method. Using only four loudspeakers we decided that both the direct and reverberant sound fields are reproduced through phantom sources. After convolving the loudspeaker signals with their respective room impulse responses (RIRs), a binaural room-in-room impulse response (BRinR) is obtained, which can again be separated into direct and reverberant components:

$$BRinR[n] = BRinR_d[n] + BRinR_r[n]. \tag{2}$$

Here, $BRinR_d[n]$ and $BRinR_r[n]$ denote the direct and reverberant components of $BRinR[n]$, reflecting the combined effects of both the simulated and playback-room acoustics.

The reproduced direct sound in the playback room further results in a direct component $BRinR_{d,d}[n]$ and a reverberant component $BRinR_{d,r}[n]$ , where these two parts are separated using a playback-room-specific separation time $T_{play}$:

$$BRinR_d[n] = BRinR_{d,d}[n] + BRinR_{d,r}[n]. \tag{3}$$

To minimize the difference between the simulated $BRIR_{ref}$ and the reproduced $BRinR$, an optimization process is performed using a Gammatone filterbank analyses and synthesis framework, which models the frequency selectivity of the human auditory system. First, within each auditory frequency band $i$, the energy of the reproduced direct sound is matched to that of the simulated $BRIR_{ref,d}[n]$ by optimizing the gain $g_d^{(i)}$:

$$ATF^{(i)}\{BRIR_{ref,d}[n]\} = (g_d^{(i)})^2 \, ATF^{(i)}\left\{BRinR_{d,d}^{(i)}[n]\right\}, \tag{4}$$

where $ATF^{(i)}\{\cdot\}$ denotes the Auditory Transfer Function, defined as the band-specific energy at the output of the $i$-th Gammatone filter. The optimized gain $g_d^{(i)}$ thus equalizes the energy of the direct sound in each auditory band between the reference and the reproduced BRinRs.

The same principle is applied to the reverberant part, where both energy and spatial diffuseness are optimized jointly. Spatial diffuseness is optimized by cross-mixing sound components, which allows to lower or increase the interaural cross-correlation to create a possibly best match between reference and room-in-room reproduction. Thus, the reverberant compensation uses frequency-dependent gains $g_r^{(i)}$ and cross-mixing coefficients matrix $CM^{(i)}$ (Fallah *et al.*, 2025):

$$\begin{aligned} ATF^{(i)}\left\{BRIR_{ref,r}[n]\right\} &= ATF^{(i)}\left\{BRinR_{d,r}[n]\right\} \\ &+ (g_r^{(i)})^2 \, ATF^{(i)}\{\ \mathbf{1}^{\top}\boldsymbol{CM}^{(i)}\boldsymbol{L}_r^{(i)}\ \}, \mathbf{1} \in \mathbb{R}^4. \end{aligned} \tag{5}$$

where $BRinR_{d,r}[n]$ is the contribution of the direct loudspeakers in reverb sound field, $CM^{(i)}$ is the cross-mixing matrix containing the frequency-dependent mixing coefficients $\alpha^{(i)}$ for front loudspeakers and $\beta^{(i)}$ for back loudspeakers, and $\boldsymbol{L}_r^{(i)}$ denotes the four loudspeakers signals that the reverberant VASs are mapped on them. The gains $g_r^{(i)}$ control the reverberant energy per band, while $CM^{(i)}$ modifies the interaural coherence (IC) to control the spatial diffuseness of the reproduced reverberant field.

This optimization yields time-invariant filters and mixing matrices that compensate for playback-room effects while preserving perceptually relevant spectral and spatial cues. The obtained coefficients $(g_d^{(i)}, g_r^{(i)}, \alpha^{(i)}, \beta^{(i)})$ are then used to construct filters for real-time rendering. The loudspeaker signals are recombined as:

$$\boldsymbol{L}^{(i)} = g_d^{(i)} \boldsymbol{L}_d^{(i)} + C^{(i)} \boldsymbol{L}_r^{(i)}, \tag{6}$$

Where $\boldsymbol{L}_d^{(i)}$and $\boldsymbol{L}_r^{(i)}$ are vetoers including  loudspeakers signals for mapped direct and reverb VASs respectively for gammatone band $i$, and $C^{(i)}$represents the composite matrix containing both cross-mixing and gain terms $(g_r^{(i)}, \alpha^{(i)}, \beta^{(i)})$. The final loudspeaker signals are resynthesized using these gains through the Gammatone filterbank. The perceptual evaluation in (Fallah *et al.*, 2025) confirmed that the ART system produces a close perceptual match to the simulated reference in both spectral balance and spatial quality.

### B. EXTENDED ART

The present study extends the ART method toward speech intelligibility testing, which requires more controlled conditions and performance metrics beyond overall sound quality. The ART framework is applied to speech intelligibility tests under the $S_0N_0$ and $S_0N_{90}$ conditions, which are standard configurations in audiological evaluation with low influence of head orientation (Warkentin *et al.*, 2024a; Warkentin *et al.*, 2024b). The block-diagram of the extended ART is depicted in FIG.1. In contrast to the previous implementation, where the direct sound was rendered using a phantom source created by panning between the two front loudspeakers, the present study employs dedicated loudspeakers for direct sound reproduction. One loudspeaker is positioned at 0° azimuth to reproduce the target speech source ($S_0$) and also the noise source ($N_0$), and another at 90° azimuth for only the noise source ($N_{90}$). This configuration avoids phantom-source generation and ensures that direct sounds are physically reproduced from their intended spatial positions. It ensures physically accurate and stable spatial reproduction of direct sounds, leading to better localization, less coloration, and potentially higher speech intelligibility.

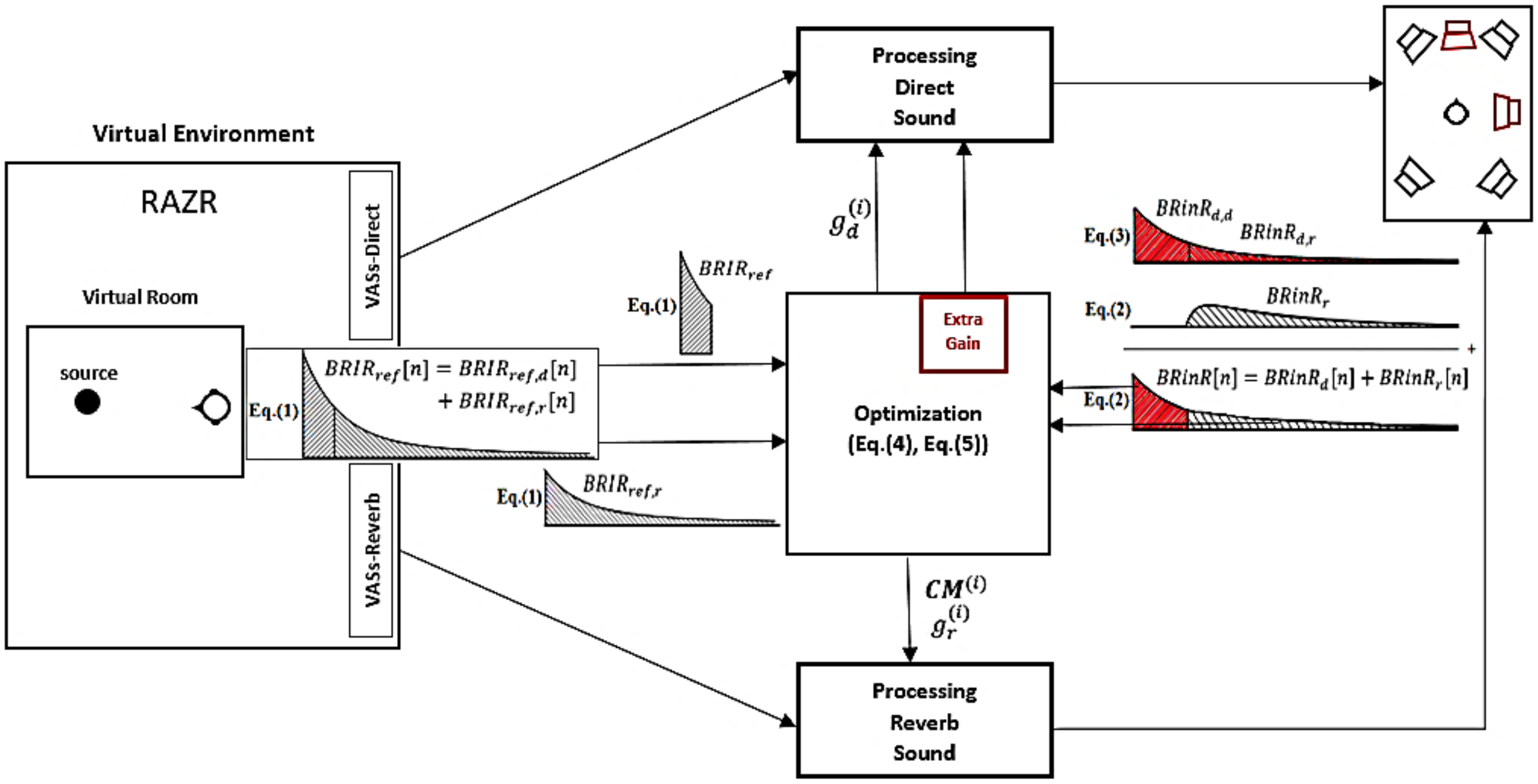

FIG.1. The block-diagram of extended ART. Unlike the previous ART setup using a phantom source panned between two front loudspeakers, direct sound is reproduced by dedicated loudspeakers at 0° ($S_0/N_0$) and 90° ($N_{90}$), providing physically stable source positions (highlighted in red in the block diagram). The reverberant component is reproduced via the same four loudspeakers as before, preserving a consistent diffuse sound field. The required BRinR impulse responses for the optimization are depicted. The optimization unit provides the required Gammatone gains and mixing coefficients for the direct- and reverb-processing blocks. To support intelligibility evaluation, the optimization is extended to additionally match the SII-weighted clarity index $C_{50,\mathrm{SII}}$. This is achieved via a 1 dB grid search over an extra frequency-independent direct-sound gain, selected to minimize the difference in $C_{50,\mathrm{SII}}$between the reference and reproduced responses.

In addition, prospectively, this will also enable a more realistic evaluation of speech intelligibility when wearing a hearing aid with directional filtering. The reverberant component is still reproduced using the same four loudspeakers, maintaining a consistent diffuse field as in the earlier setup. Altogether, the system employs six loudspeakers: two for the direct sound and four for the reverberant sound field reproduction.

To adapt the ART method for intelligibility evaluation, the optimization process is extended to include findings from previous studies on the perceptual relevance of RIRs for speech intelligibility. As shown in (Rennies *et al.*, 2011), the useful and detrimental parts of an RIR are key to understanding how reverberation affects speech perception. Based on these insights, we introduced the clarity (C50) as an additional optimization target:

$$C_{50} = \frac{\sum_{n=1}^{n_{50ms}} h^2[n]}{\sum_{n=n_{50ms}+1}^{\infty} h^2[n]}. \quad (7)$$

The $C_{50}$ value is computed for each one-third-octave frequency band and then weighted according to the Speech Intelligibility Index (SII) (ANSI, 1997). The weighted sum of the band-specific $C_{50}$ values is used to obtain the overall "$C_{50}$ index" named $C_{50}SII$:

$$C_{50}SII = \sum SII(f_{1/3oct})\, C_{50}(f_{1/3oct}) \quad (8)$$

In the first step, the $S_0$ (or $N_0$) and $N_{90}$ configurations are optimized separately, since each spatial setup represents a distinct directional source condition and corresponding pattern of reflections. In both cases, the ART filters are optimized according to the perceptual similarity criterion, which minimizes differences in energy and interaural coherence (IC) between the reproduced and reference signals. Additionally, in the extended ART, the $C_{50}SII$ is improved by adding an extra frequency-independent gain to the direct sound. The $C_{50}SII$ is evaluated assuming speech presentation from a single direction only, either 0° or 90°, corresponding to the respective spatial configuration. The separation times ($T_{ref}$, $T_{play}$) between the direct and reverberant components are determined from the empirically quality assessment as in (Fallah *et al.*, 2025). After specifying the appropriate ($T_{ref}$, $T_{play}$) pair and applying the basic ART optimization defined by equations (1)-(5), a subsequent optimization process is carried out to enhance $C_{50}SII$ in Eq.(8). For the $S_0$ condition, the left- and right-ear $C_{50}SII$ values are averaged, whereas for the $S_{90}$ condition, only the right-ear $C_{50}$ is considered since the direct sound predominantly arrives from that direction.

To optimize the $C_{50}SII$, a grid search is performed in which an additional direct-sound gain is applied and adjusted in 1 dB increments to minimize the absolute difference between the $C_{50}SII$ of the reference and that of the reproduced BRinR. The extra gain value that results in the smallest difference between the $C_{50}SII$ of the reproduced and reference responses is selected as the optimal value. The maximum additional direct-sound gain is limited to 3 dB to prevent excessive amplification of the direct sound beyond its original level. For each direct-sound gain, the reverberant cross-mixing coefficients and reverberant gains are re-optimized using Equations (1)-(5) to maintain a consistent energy balance and spatial diffuseness. This two-stage optimization enables adjustment of the direct-to-reverberant energy ratio while preserving perceptual similarity and overall sound quality. Although the direct sound deviates from the reference level, the resulting BRinR exhibits improved clarity without significant perceptible degradation.

## III. SIMULATION AND MEASUREMENTS

### A. Simulated Rooms

Two simulated rooms were generated in RAZR (Wendt *et al.*, 2014), referred to as Room 1 and Room 2. The RAZR simulations provided binaural room impulse responses (BRIRs) for headphone-based reference rendering as well as virtual acoustic sources (VASs) representing the direct and reverberant components. The reverberation times at mid frequencies (averaged over 500, 1000, and 2000 Hz) are $T_{30,\mathrm{Room1}} = 0.56$ s and $T_{30,\mathrm{Room2}} = 1.60$ s, respectively. In both simulated scenarios, the direct sound is assumed to originate from the frontal direction (0° azimuth) or from 90° azimuth, depending on the spatial configuration.

### B. Playback Room

The reproduction setup is illustrated in FIG.2. The loudspeaker configuration included six loudspeakers positioned relative to a dummy head and torso placed at the center of the room. Loudspeakers at 0° and 90° azimuth reproduced the direct sounds, while four additional loudspeakers at ±45° and ±135° azimuth were used to reproduce the reverberant field. The reverberation time of the playback room was approximately $T_{30,playback}$ = 0.41–0.45 s related to a typical office room (5m×5m×4m). The BRIR measurements were performed using Genelec 8010 loudspeakers and a KEMAR dummy head and torso (GRAS). To measure the BRIRs, excitation sweeps between 30 Hz and 22.05 kHz were played back via an RME ADI-8 QS audio interface at a sampling frequency of 44.1 kHz. Two head positions were measured: in Position 1, the head was centered at approximately 1.8 m from the front loudspeaker at 0° azimuth; in Position 2, the head was displaced 10 cm to the right. This displacement was used to evaluate the robustness of the proposed method against listener's small head movements. The head was always oriented forward during the measurements.

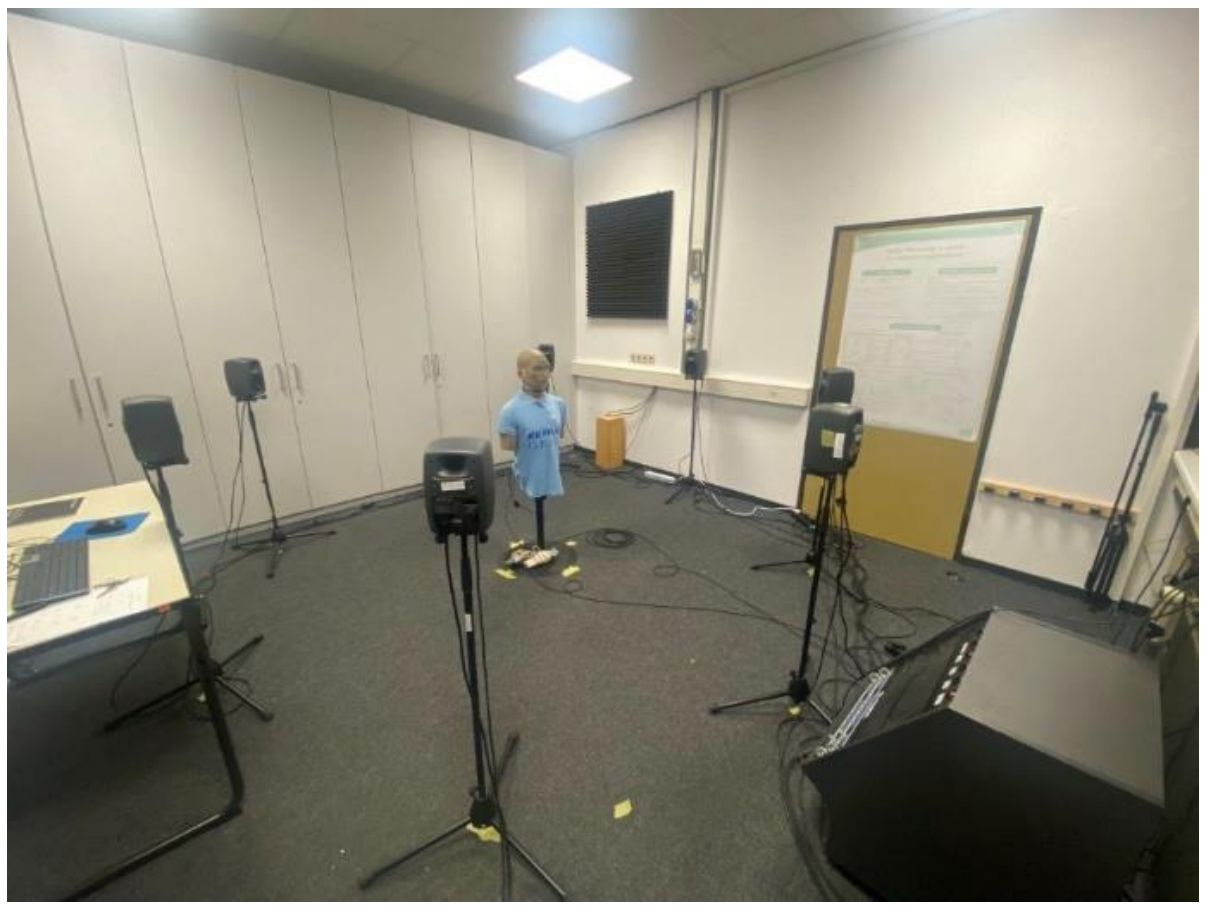

FIG.2. Playback room setup with two direct-sound loudspeakers positioned at azimuth angles of 0° and 90°, and four reverberant loudspeakers placed at ±45° and ±135°. The reverberation time ($T_{30}$) of the room was approximately 0.45 s. Binaural room impulse responses (BRIRs) were measured for all loudspeakers under different head orientations. Two head positions were used in the experiment: one with the dummy head centered and another displaced 10 cm to the right.

## IV. TECHNICAL EVALUATION OF ART OPTIMIZATION

### A. Optimization results

The optimization for the $S_0$ and $S_{90}$ scenarios was performed for both simulated rooms. In Table 1, the applied separation times (Tref – Tplay), the clarity differences ($\Delta C_{50}SII$) between the reference (Ref) and ART before applying the additional direct gain, the applied direct-sound gains, and the corresponding $\Delta C_{50}SII$ after applying these gains are summarized. In the $S_0$ condition for both rooms and the $S_{90}$ condition in Room 1, the direct sound reached the maximum allowed gain increase of 3 dB. In contrast, the $S_{90}$ condition in Room 2 required no additional amplification.

Table 1: Separation times (Tref-Tplay), additional direct-sound gains, and differences between the clarity indexes of the reference BRIR and the reproduced BRinR ($\Delta C_{50}SII$) before and after optimization.

| Scenario | ($T_{ref}$-$T_{play}$) [ms] | $\Delta C_{50}SII$ relative to Ref (no extra gain) [dB] | Extra Gain [dB] | $\Delta C_{50}SII$ relative to Ref (with extra gain) [dB] |
|---|---|---|---|---|
| S0-Room1 | (13-15) | 3.3 | 3 | 1.5 |
| S0-Room2 | (13-30) | 2.9 | 3 | 1.6 |
| S90-Room1 | (13-20) | 2.5 | 3 | 0.5 |
| S90-Room2 | (13-20) | 0.2 | 0 | 0.2 |

The 3 dB limit was imposed because complete clarity compensation at low frequencies below 1000 Hz is not feasible due to the long reverberation of the direct sound in playback room. As a result, most of the clarity improvement occurs in the mid- and high-frequency ranges, as shown in the following results.

The optimization results, including the compensation of the direct and reverberant energies, the interaural coherence (IC), and the band-dependent $C_{50}$ scores for the reproduced BRinRs in Room 1 under the $S_0$ condition, are shown in FIG.3. For this case, an additional gain of 3 dB was applied. The two upper panels of Fig. 2 display the ATFs of the reference (Ref), uncompensated (Uncomp) case, ART-processed (ART) case, and the ART-Robustness case for the left ear (upper-left panel) and the right ear (upper-right panel). For the ART-Robustness case, the dummy head was displaced by 10 cm to evaluate the robustness of the method and simulate slight listener movements. The ATFs of the reproduced BRinRs (Uncomp, ART, ART-Robustness) are compared with those of the simulated reference (Ref), which represents the ideal BRIR generated by RAZR for headphone playback.

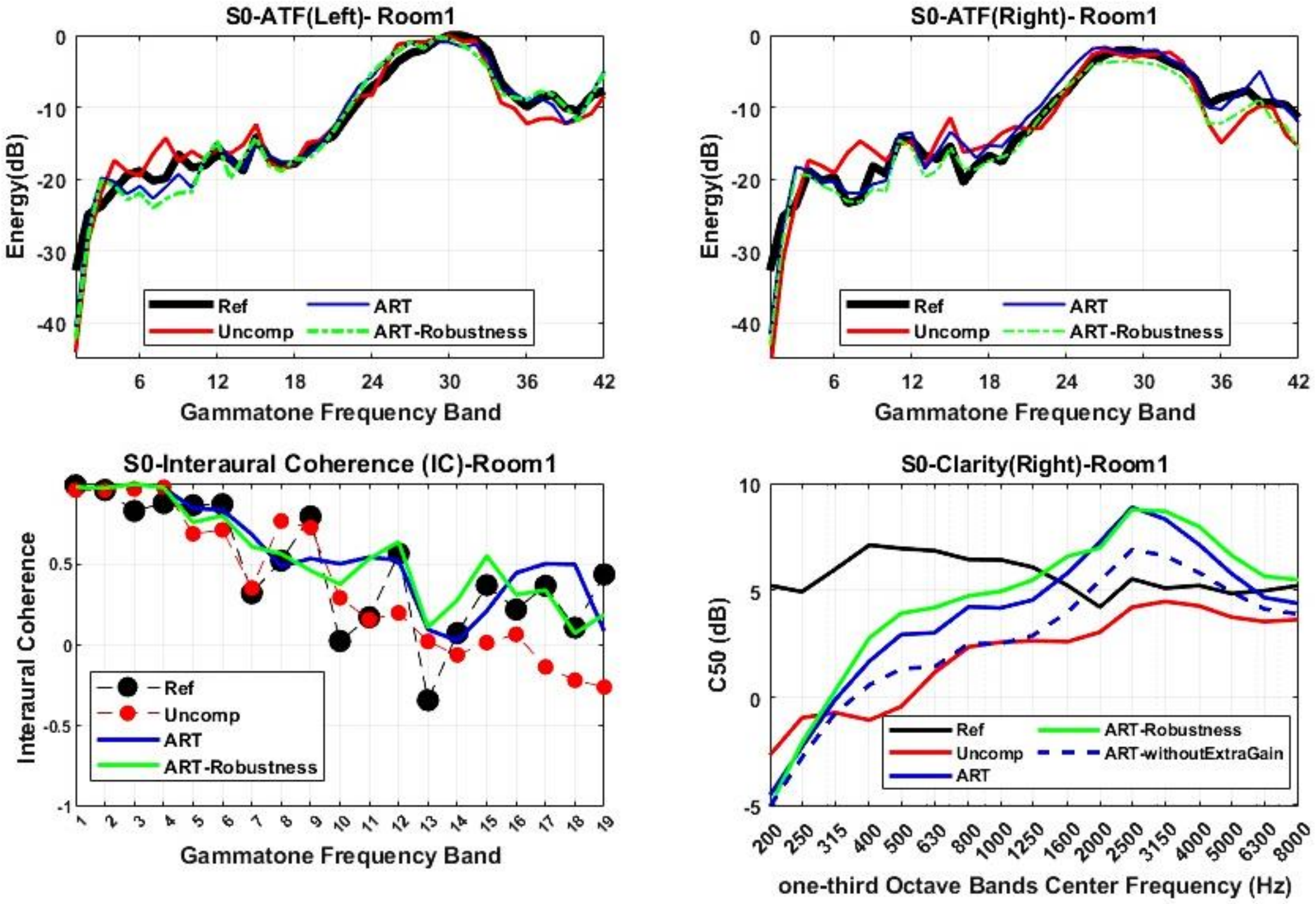

FIG.3. Optimization results for the $S_0$ scenario in Room 1 ($T_{30,Room1} = 0.56s$) with an additional 3 dB direct-sound gain. The upper panels show the ATFs of the reference (Ref), uncompensated (UnComp), ART, and ART-Robustness (10 cm head shift) cases (left and right ear results in respective upper left and right panels). The ART and ART-Robustness results exhibit reduced low-frequency energy and closer matching to the reference compared to the Uncomp case. The lower-left panel shows interaural coherence (IC) improvements across most frequency bands for ART and ART-Robustness cases. The lower-right panel presents increased average clarity ($C_{50}$) for the ART and ART-Robustness cases in comparison to the Uncomp and ART without-extra-gain cases.

In the Uncomp case, the virtual acoustic sources (VASs) from RAZR were directly mapped to the loudspeakers without any additional processing. It can be seen that, in the Uncomp condition, the reverberation of the playback room leads to an overall energy increase, particularly in the low-frequency range.

In contrast, the ATFs of the ART and ART-Robustness conditions exhibit a reduction in reproduced energy and show a closer match to the energy curve of the reference across most Gammatone bands for both ears, with nearly perfect energy matching observed for the right ear. Some mismatches between the total compensated and reference BRIRs can still be observed in the two upper panels of FIG.3, since the energy compensation was performed separately for the direct and reverberant components rather than for their combined response.

In the lower-left panel of FIG.3, the interaural coherence (IC) values of the ART-processed and uncompensated BRIRs are compared with those of the reference signal across frequency bands up to band number 19. These bands are perceptually relevant for IC, as the auditory system is particularly sensitive to changes in IC below approximately 1.5 kHz. An overall improvement in IC is observed for most bands. However, a slight reduction in the IC match occurs in a few frequency bands. This reduction can be attributed to the close reverberation times ($T_{30}$) of the simulated and playback rooms, as well as to the inherent limitations of the cross-mixing processing approach.

The lower-right panel of FIG.3 presents the $C_{50}$ for the Ref, Uncomp, and ART-processed cases across octave bands. To illustrate the improvement in clarity compared to the basic ART approach, the clarity scores obtained before applying the additional direct-sound gain (ART-withoutExtraGain) are also shown. It should be noted that only a single gain was applied to the entire BRinR, and no band-dependent gains were used to improve $C_{50}$. Such frequency-specific processing would have disrupted the energy balance and spectral coloration of the direct sound, resulting in inconsistencies with the reference signal in terms of overall energy distribution. As shown, the ART method with an additional 3 dB gain applied to the direct sound significantly increases the average clarity of the reproduced BRIRs compared with the Uncomp case and brings the results closer to the reference condition. Nevertheless, a perfect match across all frequency bands is not achieved, and the improvement is mainly reflected in the overall average $C_{50}$ score. A perfect match at lower frequencies is not achievable because reverberation is dominant in this range. This limitation also motivated the choice of an upper bound for the direct-sound gain increment. Achieving a $C_{50}$ similar to that of the reference signal at low frequencies would require a much higher gain, which would completely disrupt the energy balance between the direct and reverberant components for low- and mid-frequency ranges.

The ATFs for the left and right ears, as well as the interaural coherence (IC) and clarity index ($C_{50}$), for Room 1 under the $S_{90}$ condition are shown in FIG.4. Again, an additional gain of 3 dB was applied to the direct sound to achieve a better match with the reference (Ref) condition. Overall, a good energy match between the ART-processed BRIRs and the reference can be observed for both ears. According to the lower-left panel of FIG.4, the compensated IC does not exhibit notable improvement, particularly in the lower Gammatone bands. This can be explained by the high interaural energy differences between the left and right ear signals in the $S_{90}$ configuration, and the inherent limitations of the cross-mixing processing method. The lower-right panel again presents the $C_{50}$ for the Ref, Uncomp, and ART cases. Compared with the Uncomp condition, the ART processing results in a clear improvement in clarity across most frequency bands, as well as in the overall average $C_{50}$ score, bringing the reproduced BRIRs closer to the reference condition.

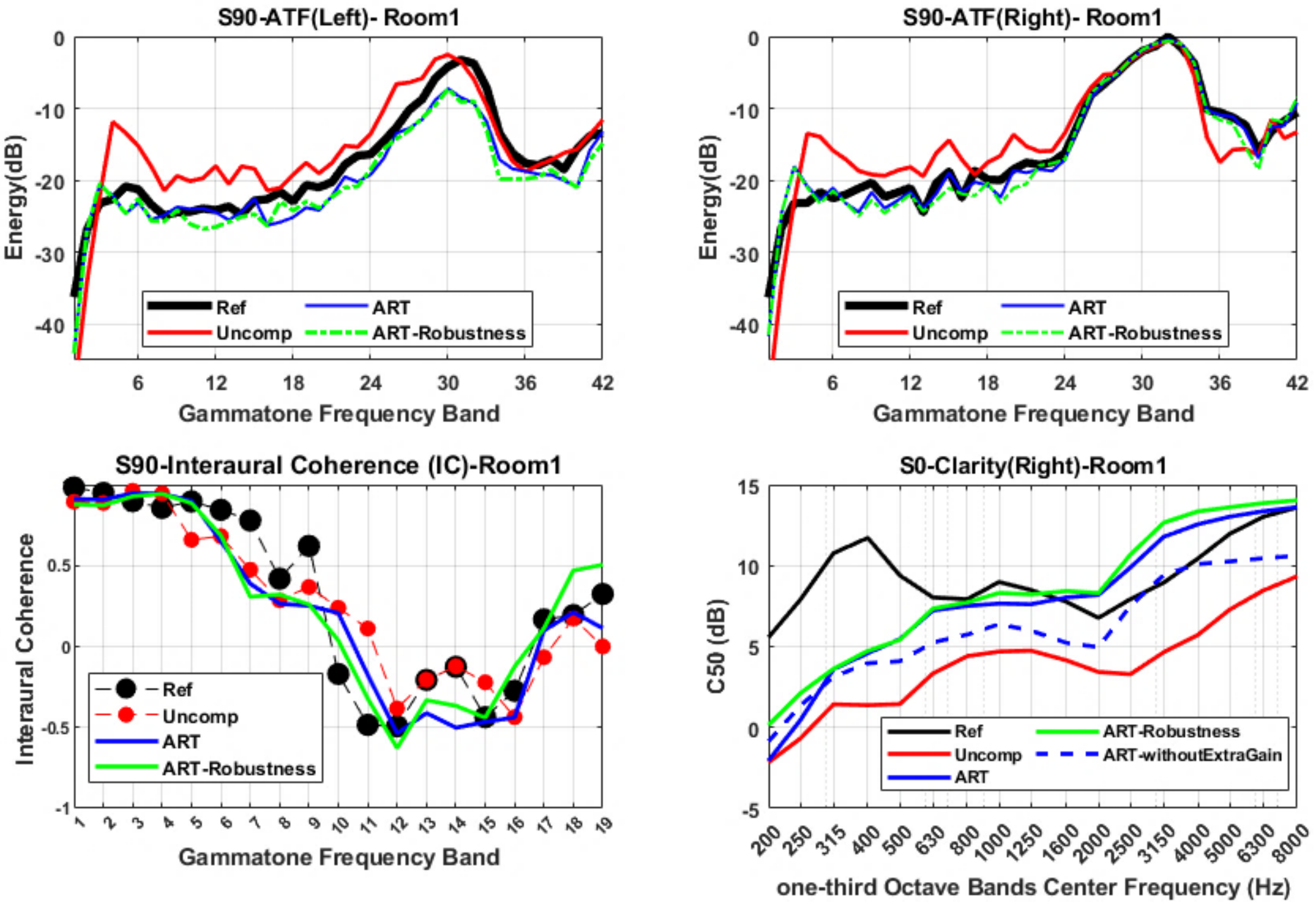

FIG.4. Similar to FIG.3, but showing results for the $S_{90}$ scenario in Room 1 with an additional 3 dB direct-sound gain. The upper panels display the ATFs of the reference (Ref), uncompensated (Uncomp), and ART-processed (ART) cases, demonstrating a good energy match for both ears. The lower-left panel shows the interaural coherence (IC), where only limited improvement is observed due to large interaural energy differences and cross-mixing limitations. The lower-right panel presents the $C_{50}$, which improves across most frequency bands and in the average score after ART processing compared with the Uncomp and ART without-Extra-Gain case.

The optimization results for Room 2 under the $S_0$ and $S_{90}$ conditions are shown in FIG.5 and FIG.6, respectively. In FIG.5, corresponding to the $S_0$ condition, a reduction in energy can be observed for the ART-processed cases, resulting in a better overall match with the reference (Ref) compared to the uncompensated (Uncomp) condition, similar to the results obtained for Room 1.

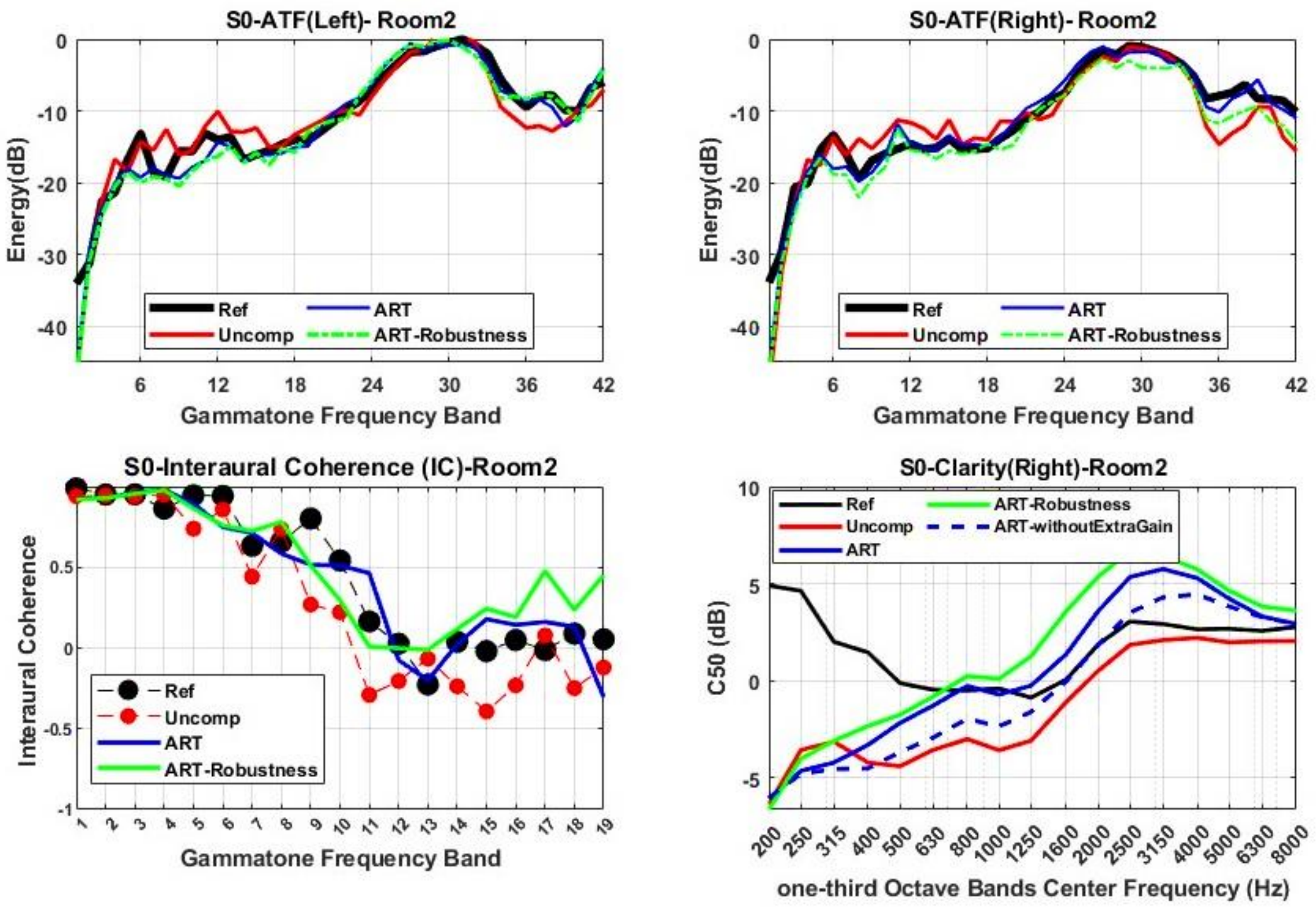

FIG.5. Optimization results for the $S_0$ scenario in Room 2 ($T_{30,Room2}$ = 1.60s), which has a longer reverberation time ($T_{30}$) than the playback room. The ART-processed cases show reduced overall energy and a closer match to the reference (Ref) compared with the uncompensated (Uncomp) condition. The lower-left panel shows improved interaural coherence (IC) compensation compared with Room 1, while the lower-right panel demonstrates increased clarity ($C_{50}$) across the frequency range and a better match to the Ref case.

As shown in the lower-left panel of FIG.5, the ART-processed case in Room 2 exhibits improved compensation for interaural coherence (IC) compared with that in Room 1. When the $T_{60}$ of the simulated room is considerably larger than that of the playback room, as in Room 2, controlling the reverberation becomes easier by adding more reverberant energy from the simulated environment.

For the $S_{90}$ condition in Room 2, as shown in FIG.6, a good energy match is achieved between the ART-processed and Ref cases.

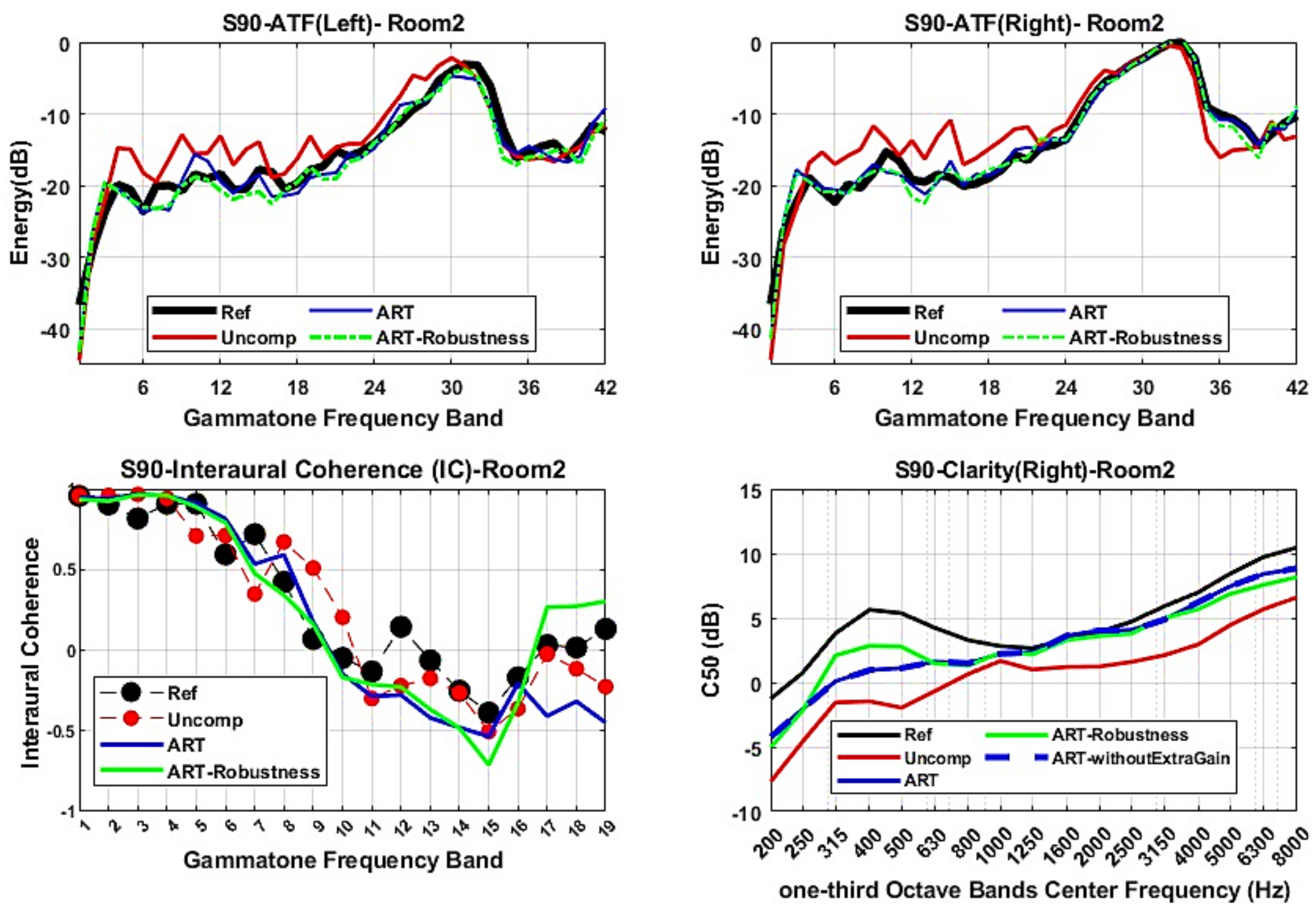

FIG.6. Similar to Fig. 4, but showing results for the $S_{90}$ scenario in Room 2. A good overall energy match is achieved between the ART-processed and reference (Ref) cases. However, as in the $S_{90}$ condition of Room 1, IC control remains limited due to interaural energy differences and cross-mixing constraints. The clarity index ($C_{50}$) shows consistent improvement for the ART-processed BRIRs compared with the Uncomp case.

Similar to Room 1, the $C_{50}$ shows improvement across the entire frequency range and a closer match to the Ref case compared with the Uncomp and ART without-Extra-Gain conditions.

was required to improve the clarity and the ART-processed BRinRs exhibits a clear improvement in the $C_{50}$ values compared with the Uncomp condition.

However, similar to the $S_{90}$ condition in Room 1 (FIG.4), controlling and improvement of IC were not achieved, due to the same interaural energy differences between left and right ears and also processing limitations. In this scenario, no additional extra gain

## V. LISTENING EXPERIMENT

### A. Stimuli

For evaluation of the proposed method, we used the German matrix test. The speech material consisted of sentences from the German Oldenburg Sentence Test (OLSA) with the male speaker (Wagener *et al.*, 1999; Wagener and Brand, 2005). The OLSA speech material consists of sentences with a fixed syntactic structure: *name, verb, number, adjective,* and *noun.* Ten alternatives are available for each word position, enabling a large number of unique sentence combinations. For speech intelligibility measurements, the sentences are organized in test lists with typically 20 sentences per list, including all 50 words of the basis list twice in different combinations. The corpus and its international counterparts (Kollmeier *et al.*, 2015) are widely used for speech intelligibility experiments due to its balanced linguistic structure and controlled level of difficulty.

For calibration, the reference BRIR and the rendered BRinR versions, including the uncompensated (Uncomp), ART, and ART-Robustness conditions, were adjusted to achieve comparable loudness levels. The OLSA speech-shaped noise, created by superimposing all sentences of the OLSA corpus, was used as the calibration signal. The levels of the reference BRIR and the reproduced BRinR were set such that, after convolution with the OLSA noise, they produced identical RMS values. After determining and applying these gain factors, each level-adjusted impulse response was convolved with all OLSA test sentences, preserving the level differences across the sentences due to intelligibility-based homogenization of the speech material.

As a fluctuating masker, the ICRA5-250 noise was employed (Wagener *et al.*, 2006). This signal belongs to the International Collegium for Rehabilitative Audiology (ICRA) family of noises, which exhibit speech-like temporal and spectral characteristics (Dreschler *et al.*, 2001). The ICRA5-250 version contains temporal gaps of up to 250 ms and has been widely used in intelligibility experiments (Holube, 2011). The combination of the OLSA speech material and ICRA5-250 noise provides a suitable framework for evaluating speech intelligibility under reverberant, fluctuating-masker conditions, where the influence of reverberation can be stronger because it temporally smears the speech signal and reduces the benefit of dip listening. The level-adjusted impulse responses were convolved with randomly selected segments of the ICRA5-250 noise to reproduce binaural noise signals.

### B. Procedure

Two spatial configurations were evaluated. The spatial configurations included $S_0N_0$, in which both the target and the noise were presented from 0° azimuth (collocated), and $S_0N_{90}$, in which the target was presented from 0° and the interferer from 90° azimuth (spatially separated).

Four rendering conditions were considered: REF (reference simulation in RAZR), Uncompensated (Uncomp) playback via loudspeakers, ART playback using ART reproduction at the sweet spot, and ART-Robustness (position-off) playback with the listener displaced 10 cm to the right, closer to the 90° loudspeaker. Each rendering condition was presented twice during the listening tests. Considering all combinations of four rendering conditions, two simulated rooms, two spatial configurations, and two repetitions per condition, resulting in a total of 32 conditions per participant, and an average duration of approximately five minutes per list (including 20 sentences each), the total measurement time per listener was about 160 minutes.

To prevent listener tiredness and maintain concentration, the 32 conditions were divided into two blocks, which were measured in separate sessions at different days. The first session evaluated the $S_0N_0$ configuration, and the second session evaluated the $S_0N_{90}$ configuration. Each block included eight conditions, and each condition was measured twice, resulting in 16 SRT measurements per block. Ten participants (4 male and 6 female), aged 19–29 years with a mean age of 24 years, and all with normal hearing, took part in the listening experiment. For half of the participants, the $S_0N_0$ configuration was presented in the first session, while for the remaining participants, the $S_0N_{90}$ configuration was presented first.

Although the OLSA sentence lists are inherently balanced for speech intelligibility, additional measures were taken to minimize potential list-dependent effects and to achieve a uniform distribution of test material across participants. Sixteen unique OLSA lists were used, corresponding to the sixteen conditions per block. To ensure a controlled experimental design, the order of sentence lists was kept constant across all listeners, while the order of test conditions was systematically rotated among participants. In this way, each condition was paired with a different list for each listener, and across all participants, every list–condition combination occurred exactly once. Ideally, a fully list-independent dataset would require sixteen participants. However, since only ten participants were available, the order of rendering conditions was arranged such that, for each simulated room, the four rendering modes (REF, Un, ART, and ART-Robustness) were separated by two list positions in the playback sequence. This arrangement ensured that eight lists were shared across rendering conditions, providing a semi-balanced distribution of test material and reducing potential list dependencies. Consequently, averaging across listeners yielded representative and largely list-independent intelligibility scores.

### C. Headphone Evaluations

The listening experiment was conducted using headphones, where the reference condition (REF) was a direct rendering of RAZR on headphones, while the remaining conditions were headphone simulated renderings such as they would occur in the playback room shown in Fig. 1, and used also for the technical evaluation. Evaluation over headphones makes the differences between experimental conditions more evident. Although in clinical applications, often loudspeaker playback would be needed, headphone presentation was chosen to avoid uncontrolled factors such as environmental noise and differences in listener height. Moreover, headphone-based evaluation provides a reliable estimate of the listener's experience with real loudspeaker reproduction, particularly regarding the coloration of the perceived sound.

The experiment was conducted with ten normal-hearing listeners in a sound-treated booth using Sennheiser HD 650 headphones. The playback chain was equalized to achieve a flat frequency response at the listener's ears using a headphone equalization filter. The reverberated noise signals were presented at a long-term average level of 65 dB SPL, corresponding to the average of the left and right ear signals in the $N_0$ condition and to the right-ear signal in the $N_{90}$ condition. The 50% speech reception thresholds ($SRT_{50}$) were measured using the AFC Toolbox in MATLAB (Ewert, 2013). The speech level of each subsequent sentence was adaptively adjusted based on the number of correctly identified words to estimate the $SRT_{50}$.

### D. Results of Listening Experiments

FIG.7 shows the $SRT_{50}$ values for ten participants obtained with the OLSA matrix test under the $S_0N_0$ condition in both rooms. In FIG.7, median values are shown together with the 25% and 75% quantiles and outliers across participants. The left and right panels in FIG.7 display the results for Room 1 and Room 2, respectively. Each of the ten listeners completed two repetitions per condition, and the two $SRT_{50}$ values were averaged to obtain one representative score per participant.

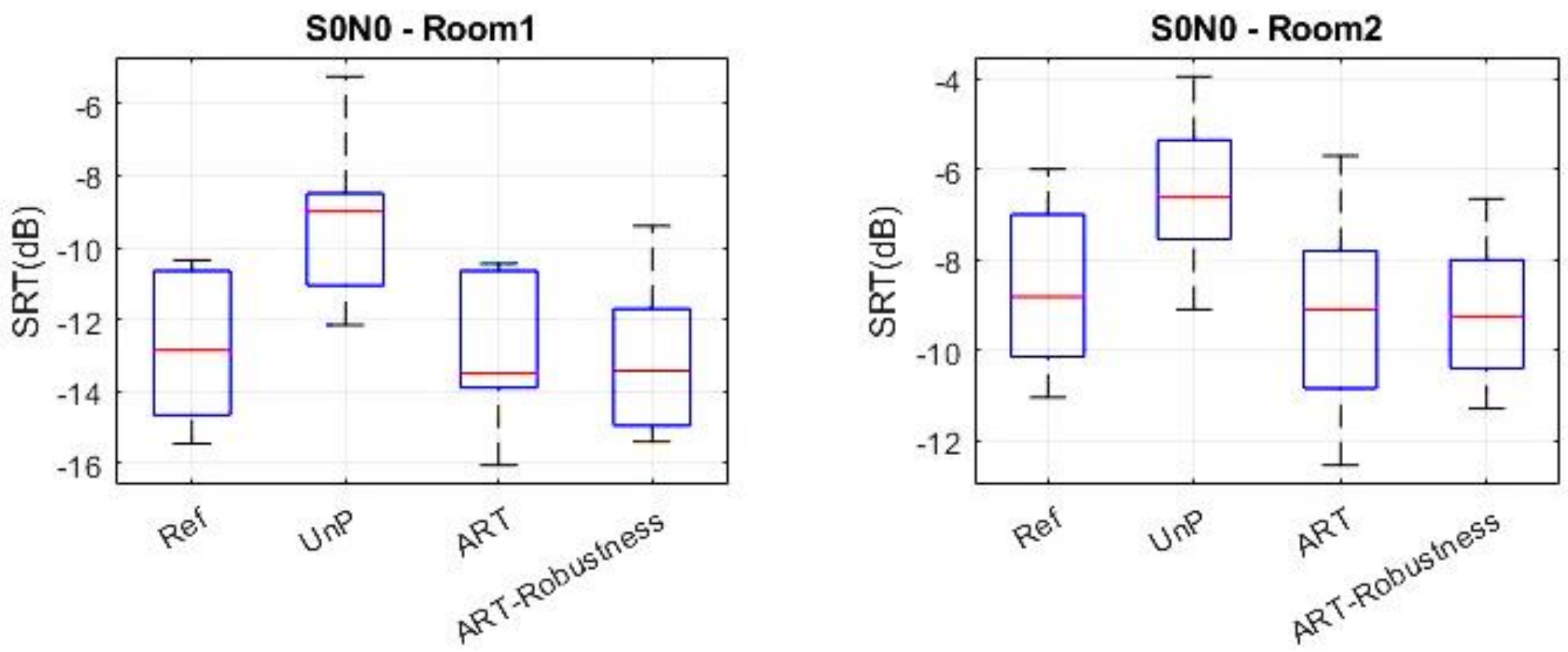

FIG.7. Speech Reception Thresholds ($SRT_{50}$) measured with the OLSA matrix test in icra5-250 noise under the $S_0N_0$ condition for ten participants for simulated Room 1 and Room 2. The two panels display the median values along with the 25th and 75th percentiles and individual outliers. Both ART and ART-Robustness produced clear improvements in SRT compared with the Uncomp condition, achieving performance levels comparable to the reference playback in both rooms.

For Room 1, a clear difference can be seen between the Uncomp condition and the other rendering conditions in both rooms (FIG.7, left). The mean $SRT_{50} \pm$ SD values are −12.87 ± 1.96 dB for Ref, −9.35 ± 2.04 dB for Uncomp, −12.78 ± 2.03 dB for ART, and −13.18 ± 1.96 dB for ART-Robustness. The effect of playback processing condition on speech reception thresholds (SRTs) was assessed using a one-way repeated-measures ANOVA. A one-way repeated-measures ANOVA revealed a significant main effect of condition on SRTs in Room 1, $F(3, 27) = 58.69$, $p = 5.90 \times 10^{-12}$, partial $\eta^2 = 0.867$. To further examine differences between individual conditions, pairwise comparisons were conducted using paired t-tests. To control for multiple comparisons, p-values were adjusted using the Holm–Bonferroni procedure. Unless stated otherwise, reported p-values correspond to Holm-corrected values. The results showed that the Uncomp condition yielded significantly higher (worse) SRTs than all other conditions (Ref vs Uncomp: $t(9) = -9.10$, $p$Holm =

0.00004; Uncomp vs ART: t(9) = 8.22, *p*Holm = 0.00007; Uncomp vs ART-Robustness: t(9) = 23.84, *p*Holm < 0.00001). In contrast, there were no significant differences among Ref, ART, and ART-Robustness (Ref vs ART: t(9) = −0.36, *p*Holm = 0.894; Ref vs ART-Robustness: t(9) = 0.85, *p*Holm = 0.894; ART vs ART-Robustness: t(9) = 1.10, *p*Holm = 0.894). These findings show that both ART-based methods substantially improved intelligibility relative to the uncompensated playback and restored performance to a level comparable to the clean reference condition. The lack of a significant difference between ART and ART-Robustness further suggests that the energy-based optimization maintains robustness against small head movements of participants during the listening tests.

A similar trend was observed in Room 2 (FIG.7, right). A one-way repeated-measures ANOVA revealed a significant main effect of playback condition on SRTs, $F(3,27) = 47.42$, $p = 6.79 \times 10^{-11}$, partial $\eta^2 = 0.840$, indicating that intelligibility varied across conditions. Holm-corrected paired t-tests confirmed that the Uncomp condition resulted in significantly poorer SRTs than the other processing conditions (Ref vs Uncomp: t(9) = −8.66, *p*Holm = 0.00005; Uncomp vs ART: t(9) = 12.24, *p*Holm < 0.00001; Uncomp vs ART-Robustness: t(9) = 9.93, *p*Holm = 0.00002). No significant differences were found among Ref, ART, and ART-Robustness (Ref vs ART: *t*(9) = 1.37, *p*Holm = 0.407; Ref vs ART-Robustness: t(9) = 2.34, *p*Holm = 0.132; ART vs ART-Robustness: t(9) = −0.06, *p*Holm = 0.954). Taken together, these results confirm that both ART variants effectively compensated for reverberation effects, producing intelligibility outcomes statistically comparable to the reference condition.

FIG.8 summarizes the results for the $S_0N_{90}$ configuration in both rooms. The figure layout matches that of Figure 6, showing medians and interquartile. Consistent with the $S_0N_0$ condition, both ART and ART-Robustness yielded substantially improved SRTs compared with the uncompensated playback, achieving values close to the reference condition in both simulated rooms.

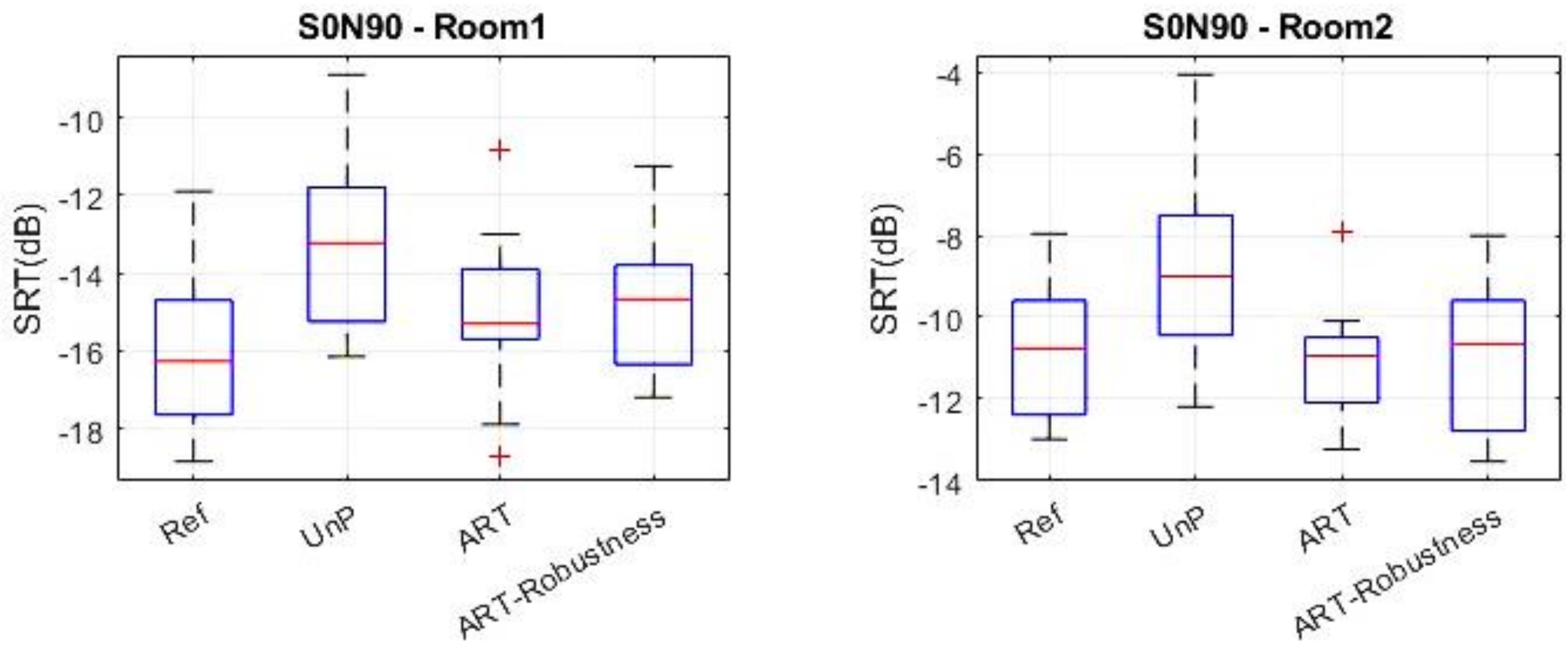

FIG.8. Speech Reception Thresholds ($SRT_{50}$) measured with the OLSA matrix test in icra5-250 noise under the $S_0N_{90}$ condition for ten participants in Room 1 and Room 2. The figure layout and analysis are identical to those in FIG.7. Similar to the $S_0N_0$ condition in FIG.7, Both ART and ART-Robustness produced clear improvements in SRT compared with the Uncomp condition.

For Room 1, the mean $SRT_{50}$ values were −15.98 ± 2.09 dB for Ref, −13.31 ± 2.35 dB for Uncomp, −15.08 ± 2.24 dB for ART, and −14.72 ± 1.78 dB for ART-Robustness. A one-way repeated-measures ANOVA revealed a significant main effect of condition on SRTs ($F(3,27) = 37.56$, $p = 8.91 \times 10^{-10}$, partial $\eta^2 = 0.807$), confirming that playback processing had a strong influence on intelligibility. Pairwise post-hoc comparisons using paired t-tests with Holm correction showed that the Uncomp condition yielded significantly poorer intelligibility than all other conditions (Ref vs Uncomp: $t(9) = -11.87$, $p\text{Holm} = 5.06 \times 10^{-6}$; Uncomp vs ART: $t(9) = 6.01$, $p\text{Holm} = 8.04 \times 10^{-4}$; Uncomp vs ART-Robustness: $t(9) = 5.22$, $p\text{Holm} = 1.66 \times 10^{-3}$). Both ART and ART-Robustness significantly improved SRTs relative to Uncomp, reaching values close to the reference condition. The difference between Ref and ART was statistically significant ($t(9) = -2.94$, $p\text{Holm} = 0.033$) with a mean SRT shift of approximately 0.9 dB. In addition, the difference between Ref and ART-Robustness was highly significant ($t(9) = -6.97$, $p\text{Holm} = 3.26 \times 10^{-4}$), indicating that the ART-Robustness condition produced higher (worse) SRTs than the reference playback. Nevertheless, both ART-based methods demonstrated strong improvements compared to the uncompensated playback. Finally, ART and ART-Robustness were not significantly different ($t(9) = -1.53$, $p$Holm = 0.160). Overall, these results indicate that both ART-based processing methods substantially enhanced speech intelligibility compared with the uncompensated playback, effectively restoring performance to near-reference levels.

For Room 2, the mean $SRT_{50}$ values were −10.77 ± 1.81 dB for Ref, −8.79 ± 2.37 dB for Uncomp, −11.03 ± 1.47 dB for ART, and −10.90 ± 1.81 dB for ART-Robustness. A repeated-measures ANOVA again showed a significant main effect of Condition ($F(3,27) = 21.05$, $p = 3.09 \times 10^{-7}$, partial $\eta^2 = 0.700$), demonstrating that intelligibility varied significantly across playback conditions. Subsequent pairwise comparisons (paired t-tests, Holm corrected) revealed that Uncomp resulted in significantly higher (worse) SRTs than Ref ($t(9) = -5.31$, $p$Holm = 0.00195), ART ($t(9) = 6.27$, $p$Holm = 0.00073), and ART-Robustness ($t(9) = 6.63$, $p$Holm = 0.00057). In contrast, no significant differences were observed between Ref and ART ($t(9) = 0.79$, $p$Holm = 1), between Ref and ART-Robustness ($t(9) = 0.43$, $p$Holm = 1), or between ART and ART-Robustness ($t(9) = -0.49$, $p$Holm = 1).

These findings confirm that both ART-based methods effectively compensated for reverberation effects in Room 2, producing intelligibility results statistically equivalent to the clean reference condition and significantly outperforming the uncompensated playback.

## VI. DISCUSSION

The present study demonstrated that the proposed six-loudspeaker Acoustic Room Transformation (ART) system successfully reproduced simulated acoustic environments in a reverberant playback room while preserving speech intelligibility close to that of the clean reference condition. Across both spatial configurations ($S_0N_0$ and $S_0N_{90}$) and in both test rooms, ART-processed playback substantially improved speech reception thresholds ($SRT_{50}$) compared with uncompensated reproduction. The consistent reduction of 2-3 dB in $SRT_{50}$ between the uncompensated (Uncomp) and compensated (ART) conditions were observed.

In the $S_0N_0$ configuration, both ART and ART-Robustness achieved $SRT_{50}$ values that were statistically indistinguishable from the reference, showing that the compensation procedure effectively neutralized the influence of the playback-room reverberation. The robustness condition, in which the listener's head was displaced by 10 cm, produced nearly identical intelligibility performance, indicating that the optimization generalized well to small spatial variations. This finding is important for practical applications, as perfect listener positioning cannot always be guaranteed in real listening tests or clinical environments. In the $S_0N_{90}$ configuration, where spatial separation typically yields a binaural release from masking (Bronkhorst and Plomp, 1992) , the ART method maintained this advantage. Both ART and ART-Robustness produced $SRT_{50}$ scores within 1 dB of the reference in both rooms, confirming that the ART framework preserves interaural cues essential for spatial unmasking.

The success of the approach is due to its perceptually grounded optimization process. By decomposing the simulated and playback responses into direct and reverberant components and matching them via Gammatone-filterbank analysis, the algorithm controls the effective direct-to-reverberant energy ratio (D/R) without altering spectral coloration. The additional clarity optimization using the SII-weighted $C_{50}$ ensured perceptual relevance across frequency bands, while the 3 dB limit on additional direct-sound gain prevented over-amplification and preserved a natural balance between early and late energy. This limitation is particularly justified at low frequencies, where long decay times make full compensation of reverberation physically infeasible (Toole, 2006; Kuttruff, 2016). The findings align with earlier studies showing that enhancing the direct component or early reflections can improve speech intelligibility in reverberant environments (Buchholz and Favrot, 2009; Ahrens et al., 2019; Prodi et al., 2022).

The fact that both ART and ART-Robustness achieved near-reference SRTs in two acoustically distinct rooms suggests that the approach is robust to variations in reverberation time and energy distribution. In Room 1, which exhibited a shorter $T_{60}$, the small but significant difference between Ref and ART (≈0.9 dB) may reflect a slight under-compensation of the direct field. In the more reverberant Room 2, however, no significant differences were observed among Ref, ART, and ART-Robustness, implying that the optimization adapts effectively when late reflections dominate the room response.

Despite these promising results, several directions for improvement and future work remain. First, in the current implementation, the same cross-mixing coefficients were used for both ears. Especially in the $S_0N_{90}$ condition, where interaural coherence plays a key role in spatial unmasking, future research could introduce ear-specific mixing coefficients to more precisely control interaural coherence and binaural cues. The main cost of this modification is the increased computational load introduced by the grid-search algorithm. Second, the clarity enhancement in this study was achieved by increasing the direct-sound energy, which may not always be the perceptually optimal strategy. An alternative could be to attenuate the reverberant component instead, achieving the desired $C_{50}$ improvement through reverberation reduction rather than amplification of the direct field. Third, using more directional loudspeakers for reproducing the direct sound may further improve spatial precision and minimize undesired acoustic overlap between direct and reverberant components. Such an approach would make the system more adaptable to rooms with strong early reflections.

Beyond these technical extensions, future studies should also evaluate the method with hearing-impaired listeners using hearing aids, incorporate dynamic listener movements, and validate the approach in real loudspeaker environments rather than headphone-based playback. In that context, it will also be important to measure SRTs employing the ART system to compensate for different playback rooms using the listeners.

Overall, the experimental results confirm that perceptually optimized ART processing effectively compensates for the acoustic coloration and intelligibility degradation introduced by real playback rooms. The findings demonstrate that even with a compact loudspeaker setup, it is possible to achieve perceptually equivalent performance to that of ideal, simulated reference conditions. This makes the ART system a practical and flexible tool for speech intelligibility and listening-effort studies in realistic environments.

## CONCLUSION

This study introduced and evaluated an extended Acoustic Room Transformation (ART) system designed to reproduce simulated acoustic environments in reverberant playback rooms for performing speech intelligibility tests. The method separates direct and reverberant sound components, optimizes their balance using a perceptually motivated framework based on Gammatone-filterbank analysis, and enhanced clarity through an SII-weighted $C_{50}$ optimization constrained by a 3 dB direct-sound gain limit.

Listening tests with ten normal-hearing participants demonstrated that both the ART and ART-Robustness conditions significantly improved speech intelligibility compared with uncompensated playback. Across two playback rooms and two spatial configurations ($S_0N_0$ and $S_0N_{90}$), the resulting $SRT_{50}$ scores closely matched those of the simulated reference, showing that the approach effectively compensates for reverberation and preserves key spatial cues. Importantly, the robustness evaluation condition preserved performance despite deviations from the sweet spot, confirming the practical stability of the system.

Overall, the results indicate that perceptually guided reproduction using a small loudspeaker array can achieve speech intelligibility levels equivalent to those obtained under ideal reference conditions. This finding has direct implications for speech and hearing research, clinical testing, and the design of intelligibility evaluation setups in non-anechoic environments.

Future work will focus on refining the method by introducing ear-specific mixing coefficients to better control interaural coherence in spatially separated conditions, reducing reverberant energy instead of merely amplifying the direct sound to enhance clarity, and employing directional loudspeakers to improve spatial precision of direct-sound reproduction. Evaluations with hearing-impaired listeners and dynamic listener movements will also be essential to fully establish the ecological and clinical applicability of the ART framework.

**ACKNOWLEDGMENTS**

This study was funded by the Deutsche Forschungsgemeinschaft (DFG, German Research Foundation) – Project ID: 352015383 – SFB 1330, C2/C4. The AI-based tool ChatGPT5 (OpenAI) was used to assist with language editing during the preparation of this manuscript.

**Author Declarations**

**Conflict of Interest**

The authors declare no conflicts of interest related to the content of this article.

**Ethics Approval**

The listening tests conducted in this study were approved by the appropriate ethics committee at Carl von Ossietzky University of Oldenburg, and all participants provided informed consent in accordance with institutional guidelines.

**Data Availability**

The data that support the findings of this study are available from the corresponding author upon request. Audio examples, simulation scripts, and evaluation materials used in the listening tests can be shared for academic research purposes.

## REFERENCES

Ahrens, A., Marschall, M., and Dau, T. (**2019**). "Measuring and modeling speech intelligibility in real and loudspeaker-based virtual sound environments," Hearing research **377**, 307-317.

ANSI, S.-. (**1997**). "Methods for calculation of the speech intelligibility index," American National Standards Institute, New York.

Badajoz-Davila, J., Buchholz, J. M., and Van-Hoesel, R. (**2020**). "Effect of noise and reverberation on speech intelligibility for cochlear implant recipients in realistic sound environments," The Journal of the Acoustical Society of America **147**, 3538-3549.

Bronkhorst, A., and Plomp, R. (**1992**). "Effect of multiple speechlike maskers on binaural speech recognition in normal and impaired hearing," The Journal of the Acoustical Society of America **92**, 3132-3139.

Buchholz, J. M., and Favrot, S. (**2009**). "A loudspeaker-based room auralisation system for auditory perception research," in *NAG/DAGA 2009 International Conference on Acoustics*, pp. 1295-1298.

Cueille, R., Lavandier, M., and Grimault, N. (**2022**). "Effects of reverberation on speech intelligibility in noise for hearing-impaired listeners," Royal Society Open Science **9**, 210342.

Dong, H.-Y., and Lee, C.-M. (**2018**). "Speech intelligibility improvement in noisy reverberant environments based on speech enhancement and inverse filtering," EURASIP Journal on Audio, Speech, and Music Processing **2018**, 3.

Dreschler, W. A., Verschuure, H., Ludvigsen, C., and Westermann, S. (**2001**). "ICRA Noises: Artificial Noise Signals with Speech-like Spectral and Temporal Properties for Hearing Instrument Assessment: Ruidos ICRA: Señates de ruido artificial con espectro similar al habla y propiedades temporales para pruebas de instrumentos auditivos," Audiology **40**, 148-157.

Ewert, S. D. (**2013**). "AFC—A modular framework for running psychoacoustic experiments and computational perception models," in *Proceedings of the international conference on acoustics AIA-DAGA* (Deutsche Gesellschaft für Akustik eV Berlin), pp. 1326-1329.

Fallah, A., Gündert, S., Kirsch, C., Nakamura, S., and van de Par, S. (**2023**). "Reproduction of simulated acoustic scenes with limited number of loudspeakers in a reverberant room," in *Forum Acousticum 2023* (Turin Italy).

Fallah, A., Nakamura, S., Gündert, S., Kirsch, C., and van de Par, S. (**2025**). "Reproduction of simulated acoustic scenes for clinical and hearing research in a reverberant room using four horizontal loudspeakers," Acta Acustica **9**, 64.

Favrot, S., and Buchholz, J. M. (**2010**). "LoRA: A loudspeaker-based room auralization system," Acta acustica united with Acustica **96**, 364-375.

Gerken, M., Hohmann, V., and Grimm, G. (**2024**). "Comparison of 2D and 3D multichannel audio rendering methods for hearing research applications using technical and perceptual measures," Acta Acustica **8**, 17.

Guastamacchia, A., Riente, F., Shtrepi, L., Puglisi, G. E., Pellerey, F., and Astolfi, A. (**2024**). "Speech intelligibility in reverberation based on audio-visual scenes recordings reproduced in a 3D virtual environment," Building and Environment **258**, 111554.

Haeussler, A., and van de Par, S. (**2019**). "Crispness, speech intelligibility, and coloration of reverberant recordings played back in another reverberant room (Room-In-Room)," The Journal of the Acoustical Society of America **145**, 931-944.

Hamdan, E. C., and Fletcher, M. D. (**2022**). "A Compact Two-Loudspeaker Virtual Sound Reproduction System for Clinical Testing of Spatial Hearing With Hearing-Assistive Devices," Frontiers in Neuroscience **15**, 725127.

Holube, I. (**2011**). "Speech intelligibility in fluctuating maskers," in *Proceedings of the international symposium on auditory and audiological research*, pp. 57-64.

Kates, J. M., Lavandier, M., Muralimanohar, R. K., Lundberg, E. M., and Arehart, K. H. (**2025**). "Binaural speech intelligibility for combinations of noise, reverberation, and hearing-aid signal processing," PloS one **20**, e0317266.

Kollmeier, B., Warzybok, A., Hochmuth, S., Zokoll, M. A., Uslar, V., Brand, T., and Wagener, K. C. (**2015**). "The multilingual matrix test: Principles, applications, and comparison across languages: A review," International journal of audiology **54**, 3-16.

Kuttruff, H. (**2016**). *Room acoustics* (Crc Press).

Mansour, N., Marschall, M., May, T., Westermann, A., and Dau, T. (**2021**). "Speech intelligibility in a realistic virtual sound environment," The Journal of the Acoustical Society of America **149**, 2791-2801.

Ooster, J., Krueger, M., Bach, J.-H., Wagener, K. C., Kollmeier, B., and Meyer, B. T. (**2020**). "Speech audiometry at home: Automated listening tests via smart speakers with normal-hearing and hearing-impaired listeners," Trends in Hearing **24**, 2331216520970011.

Prodi, N., Pellegatti, M., and Visentin, C. (**2022**). "Effects of type of early reflection, clarity of speech, reverberation and diffuse noise on the spatial perception of a speech source and its intelligibility," The Journal of the Acoustical Society of America **151**, 3522-3534.

Puglisi, G. E., Warzybok, A., Astolfi, A., and Kollmeier, B. (**2021**). "Effect of reverberation and noise type on speech intelligibility in real complex acoustic scenarios," Building and Environment **204**, 108137.

Pulkki, V. (**1997**). "Virtual sound source positioning using vector base amplitude panning," Journal of the audio engineering society **45**, 456-466.

Reinhart, P. N., and Souza, P. E. (**2016**). "Intelligibility and clarity of reverberant speech: Effects of wide dynamic range compression release time and working memory," Journal of Speech, Language, and Hearing Research **59**, 1543-1554.

Rennies, J., Brand, T., and Kollmeier, B. (**2011**). "Prediction of the influence of reverberation on binaural speech intelligibility in noise and in quiet," The Journal of the Acoustical Society of America **130**, 2999-3012.

Schütze, J., Kirsch, C., Kollmeier, B., and Ewert, S. D. (**2025**). "Comparison of speech intelligibility in a real and virtual living room using loudspeaker and headphone presentations," Acta Acustica **9**, 6.

Toole, F. E. (**2006**). "Loudspeakers and rooms for sound reproduction—A scientific review," Journal of the Audio Engineering Society **54**, 451-476.

Wagener, K., Kühnel, V., and Kollmeier, B. (**1999**). "Development and evaluation of a German sentence test I: Design of the Oldenburg sentence test," Zeitschrift Fur Audiologie **38**, 4-15.

Wagener, K. C., and Brand, T. (**2005**). "Sentence intelligibility in noise for listeners with normal hearing and hearing impairment: Influence of measurement procedure and masking parameters La inteligibilidad de frases en silencio para sujetos con audición normal y con hipoacusia: la influencia del procedimiento de medición y de los parámetros de enmascaramiento," International journal of audiology **44**, 144-156.

Wagener, K. C., Brand, T., and Kollmeier, B. (**2006**). "The role of silent intervals for sentence intelligibility in fluctuating noise in hearing-impaired listeners: El papel de los intervalos de silencio para la inteligibilidad de frases en medio de ruido fluctuante en sujetos hipoacùsicos," International Journal of Audiology **45**, 26-33.

Warkentin, L., Denk, F., Winkler, A., Sankowsky-Rothe, T., Blau, M., Holube, I., and Husstedt, H. (**2024a**). "Effect of room acoustic properties and head orientation on practical speech-in-noise measurements for various spatial configurations," Acta Acustica **8**, 24.

Warkentin, L., Holube, I., Winkler, A., Denk, F., Sankowsky-Rothe, T., Blau, M., and Husstedt, H. (**2024b**). "Verwendung von Sprachtests im Freifeld in Deutschland."

Wendt, T., Van De Par, S., and Ewert, S. D. (**2014**). "A computationally-efficient and perceptually-plausible algorithm for binaural room impulse response simulation," Journal of the Audio Engineering Society **62**, 748-766.

Winkler, A., Warkentin, L., Denk, F., Husstedt, H., Sankowksy-Rothe, T., Blau, M., and Holube, I. (**2025**). "Reference Speech-recognition curves for a German monosyllabic test in noise: effects of loudspeaker configuration and room acoustics," International Journal of Audiology **64**, 695-704.

Xia, J., Xu, B., Pentony, S., Xu, J., and Swaminathan, J. (**2018**). "Effects of reverberation and noise on speech intelligibility in normal-hearing and aided hearing-impaired listeners," The Journal of the Acoustical Society of America **143**, 1523-1533.

Zucatelli, G., and Coelho, R. (**2021**). "Harmonic and non-harmonic based noisy reverberant speech enhancement in time domain," arXiv preprint arXiv:2112.04949.